\documentclass[twocolumn,showpacs,amsmath,amssymb,showkeys,floatfix,prl]{revtex4}
\usepackage{amsfonts,amsmath}
\usepackage{graphicx}
\usepackage{dcolumn}% Align table columns on decimal point
\usepackage{bm}% bold math

\begin{document}

\title{Measuring the mass of a sterile neutrino with a very short baseline 
reactor experiment}

\author{D. C. Latimer$^1$, J. Escamilla$^2$  and D. J. Ernst$^2$}

\affiliation{
$^{1}$ School of Liberal Arts and Sciences, Cumberland University, Lebanon,
Tennessee 37087}
\affiliation{$^{2}$ Department of Physics and Astronomy, Vanderbilt University, 
Nashville, Tennessee 37235
}

\date{\today}

\begin{abstract}
An analysis of the world's neutrino oscillation data, including sterile neutrinos, (M. Sorel, C. M. Conrad,
and M. H. Shaevitz, Phys. Rev. D {\bf 70}, 073004) found a peak in the allowed region 
at a mass-squared difference $\Delta m^2\cong 0.9$ eV$^2$. We trace its origin 
to harmonic oscillations in the electron survival probability ${\mathcal P}_{ee}$ as a 
function of $L/E$, the ratio of baseline to neutrino energy, as measured in the near detector of the Bugey experiment.
We find a second occurrence for $\Delta m^2\cong 1.9$ 
eV$^2$. We point out 
that the phenomenon of harmonic oscillations of ${\mathcal P}_{ee}$ as a function of $L/E$, 
as seen in the Bugey experiment, can be used to measure the mass-squared difference associated with a sterile 
neutrino in the range from a fraction of an eV$^2$ to several eV$^2$ (compatible with that indicated by 
the LSND experiment), as well as measure the amount of electron-sterile neutrino mixing.
We observe that the experiment is independent, to lowest order, of the size of the reactor and suggest the 
possibility of a small reactor with a detector sitting at a very short baseline.
\end{abstract}



\maketitle

Present data demonstrate that neutrinos change their flavor while propagating in vacuum and 
through matter. The evidence comes from 
solar neutrino experiments \cite{solar}, a long baseline reactor experiment \cite{kaml}, atmospheric experiments \cite{atmo}, 
and a long baseline 
accelerator experiment \cite{k2k}. These experiments, together with the constraint 
imposed by the CHOOZ  reactor experiment \cite{choo}, provide a quantitative \cite{glf06}
determination of the mixing parameters and mass-squared differences for three neutrino oscillations. Moreover, as the data 
becomes ever more precise, alternative explanations of the data are 
continually being ruled out \cite{mcgg04}.

The lone datum that does not fit into the scenario of three neutrino mixing is the appearance result from the  
LSND experiment \cite{lsnd}.  An oscillation explanation of this result requires a neutrino mass-squared difference of at least $10^{-1}$ eV$^2$ while 
the world's remaining data is 
compatible with two mass-squared differences of the order of $8 \times 10^{-5}$ eV$^2$  and $2 \times 10^{-3}$ eV$^2$. 
The addition of a sterile neutrino or neutrinos \cite{sterilenu} has been proposed in an attempt to incorporate LSND into an analysis that would
be consistent with the world's data. Other physical mechanisms have also been proposed as possible explanations \cite{other}. 

Restricting our discussion to sterile neutrinos (with CP and CPT conserved), the simplest extension is the inclusion of a single sterile neutrino 
\cite{fournu}. Such models fall into one of two classes.  The 3+1 scheme adds the fourth neutrino whose mass separation is much larger than the other three.  
In the 2+2 scheme, the LSND mass-squared difference separates two pairs of neutrinos 
with the smaller mass-squared differences.
Current analyses indicate that neither provides a compelling explanation of the data \cite{fournustatus,sore}.
As four neutrinos do not seem to be sufficient, five neutrino scenarios have been investigated \cite{sore,fivenu}.  

\begin{figure}[ht]
\begin{center}
\includegraphics*[width=3in]{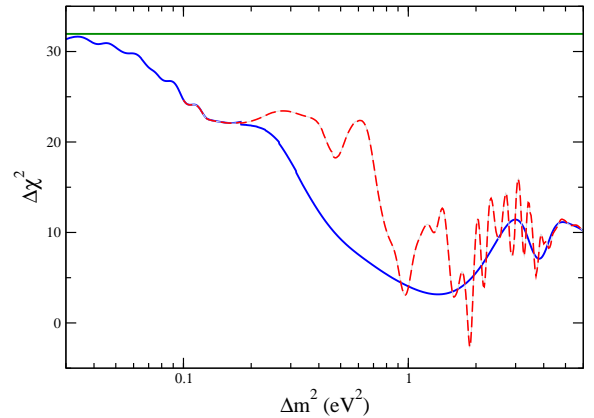}
\caption{The value of $\Delta \chi^2=\chi^2-\chi^2_{min}$ versus $\Delta m^2$.
The solid (green) straight line is the result for a three-neutrino fit to data from Refs. \protect\cite{solar,kaml,atmo,k2k,choo,lsnd,karm}. The solid (blue) 
curve is the result of a four-neutrino fit to this data, and the dashed (red) curve is the 
result if, in addition, we add Bugey \protect\cite{buge}. The zero for the vertical scale 
($\chi^2_{min}$) is arbitrary.}
\label{fig1}
\end{center}
\end{figure}

Here we assume the existence one sterile neutrino in a 3+1 scheme and
propose a new way to experimentally measure the associated mass-squared difference and mixing angle. 
For simplicity, we introduce only one sterile neutrino; however, our methods can be easily
generalized to include additional sterile neutrinos.
For the baselines and energies of 
interest here,
only the mass-squared differences involving the extra neutrino will contribute to the oscillations.
The mass of the three usual neutrinos can be taken as degenerate so that oscillations will be governed by only one mass-squared difference $\Delta m^2$.

In Fig.~4 of Ref.~\cite{sore}, there is a narrow peak in the allowed region which occurs at 
$\Delta m^2\cong 0.9$ eV$^2$. We can trace this peak to the Bugey 
reactor experiment \cite{buge}, which is an electron anti-neutrino disappearance experiment with detector baselines of 
15, 40, and 95 m.  
We construct a model of the aforementioned neutrino experiments.  This analysis \cite{us} produces mixing angles for three neutrino oscillations that are 
very similar to those of Ref.~\cite{glf06}.
In Fig.~\ref{fig1}, we present $\chi^2$ versus the mass-squared difference, $\Delta m^2$, for three cases. 

The horizontal solid (green) line 
is the result of a three-neutrino analysis of the data from Refs.~\cite{solar,kaml,atmo,k2k,choo}, LSND \cite{lsnd}, and KARMEN \cite{karm}. 
The $\chi^2$ contains the no oscillation contribution from LSND and KARMEN.
 By 
definition the three neutrino results do not depend on $\Delta m^2$ yielding a straight line. 
The solid (blue) curve is the result of a 3+1 analysis. As expected, the nonzero LSND data utilizes $\Delta m^2$ and the additional mixing angles $\theta_{14}$ and $\theta_{24}$  (the results are essentially independent of $\theta_{34}$) to lower the $\chi^2$. For $\Delta m^2<0.03$ 
eV$^2$ the fourth neutrino does not contribute to the 
LSND or KARMEN experiment,
and the $\chi^2$ reverts to the three neutrino result as it must. 
The exact values of these two curves are irrelevant for the 
discussion at 
hand. We include them to provide a reasonable background
for the dashed curve in which we add the Bugey experiment \cite{buge} to the previous analysis, following exactly the analysis in Ref.~\cite{buge}.  
For $\Delta m^2 < 0.2$ eV$^2$,  the dashed curve merges with the solid curve as Bugey does not 
contribute in this region.   Beyond this, the dashed
curve contains fluctuations.  This is because the parameters are in a region that runs along the edge of the 
Bugey excluded region where the $\chi^2$ is not smooth. This phenomena is also present in Ref.~\cite{sore}. These curves do {\it not} indicate the 
existence of a sterile neutrino. The addition of the CDHS \cite{cdhs}, CCFR84 \cite{ccfr}, and NOMAD \cite{noma} experiments were found in 
Ref.~\cite{sore} to largely offset the LSND indication of a sterile neutrino.

The first important feature of the dashed curve is the narrow dip near 0.9 eV$^2$. This corresponds to the peak in the allowed region found in
Fig.~4 of Ref.~\cite{sore}. Note that we also find a second  narrow dip near 1.9 eV$^2$. This larger mass-squared difference is also present
in Ref.~\cite{sore};  however, 
in their analysis, the inclusion of other null result short baseline experiments \cite{cdhs,ccfr,noma} almost completely suppresses this dip's significance. 
Clearly the narrow dip in the $\chi^2$ here and the narrow peak in the allowed region in
Ref.~\cite{sore} originates from the Bugey experiment.

\begin{figure}
\begin{center}
\includegraphics*[width=3in]{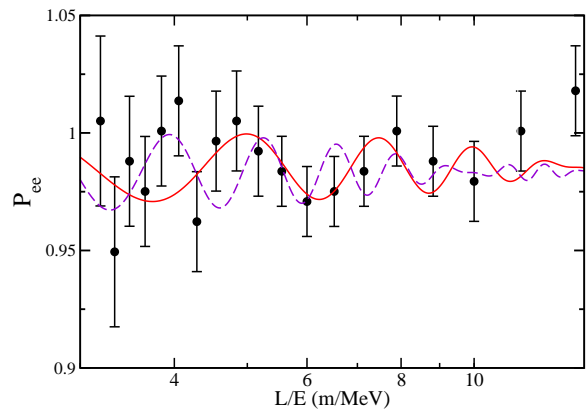}
\caption{The electron survival probability ${\mathcal P}_{ee}$ versus $L/E$. The 
data are from the near detector, $L=15$ m, of the 
Bugey \cite{buge} experiment. The solid (red) curve results from a 
sterile neutrino associated with $\Delta m^2 = 0.9$ eV$^2$ 
and the dashed (violet) curve 
results from $\Delta m^2 = 1.9$ eV$^2$.}
\label{fig2}
\end{center}
\end{figure}

To understand the source of this dip, we turn to the {\it in vacuo} neutrino oscillation probability.  Using a standard extension of the MNS mixing matrix, the electron neutrino survival probability at these short baselines is approximately
%The probability that a neutrino of flavor $\alpha$ and energy $E$ (in MeV) will be detected as a neutrino of flavor $\beta$ a 
%distance $L$ (in m) from the source is 
%\begin{equation}
%\mathcal{P}_{\alpha  \beta}(L/E)
%= \delta_{\alpha \beta}
%-4 \sum^{3+1}_{\genfrac{}{}{0pt}{}{j >
%k}{j,k=1}} U_{\alpha j} U_{\alpha k} U_{\beta k} 
%U_{\beta j} \sin^2 \phi_{jk}
%\label{oscform}
%\end{equation}
%\begin{equation}
%\mathcal{P}_{ee}(L/E) \cong 1-4(1-U_{44}^2)U_{44}^2 \sin^2 \phi.
%\end{equation}
%with $\phi = 1.27 \Delta m^2 L/ E$.  Using the standard parameterization of $U$, one can define  a mixing angle $\theta_{14}$ such that $U_{44} = \sin \theta_{14}$; this yields essentially a two neutrino oscillation probability at short baselines
\begin{equation}
\mathcal{P}_{ee}(L/E) \cong 1-\sin^2(2\,\theta_{14}) \sin^2 \phi,
\end{equation}
with $\phi = 1.27 \Delta m^2 L/ E$ where $\Delta m^2$ is in eV$^2$, the baseline $L$ is in m, and the energy $E$ is in MeV.
We can use this to determine which data in the Bugey experiment yield the dips in 
$\chi^2$.  Fig.~\ref{fig2} contains a plot 
of ${\mathcal P}_{ee}$ versus $L/E$ from the Bugey near detector located at $L=15$ m. The solid (red) curve is for the best 
fit
parameters with a mass-squared difference of $\Delta m^2 = 0.9$ eV$^2$ and the dashed (violet) curve
for $\Delta m^2 = 1.9$ eV$^2$. Although the statistics are poor, one can see that there is an harmonic oscillation in the data at frequencies which 
generate the dips in the $\chi^2$. The
dips correspond roughly to a change in $\chi^2$ of ten. Whether these dips are real or statistical 
fluctuations is not the question. The point is that with improved statistics the existence of a sterile neutrino with $\Delta m^2$ in 
the appropriate range would produce a measurable narrow minimum in the  $\chi^2$.
The Bugey experiment is an existence proof for the validity of such an experiment.

For a reactor, the neutrino spectrum and 
technology set the range of detectable neutrino energies to lie between approximately 1 to 5 MeV.
At a baseline of 15 m, the ratio $L/E$ ranges from 
3 to 15 m/MeV. Oscillations have undergone one cycle when $\phi=\pi$.  For a $\Delta m^2$ of 0.9 (1.9) eV$^2$, the oscillation length is
2.7 (1.3) m/MeV which
results in approximately 5 (10) cycles in the allowed range.

The amplitude of the oscillations for the Bugey data is about 1.25\%. Twice this (peak to trough) is somewhat less than the average
error bar thus giving the low statistical significance to these dips. If the statistical error bars were one fourth this (sixteen times the data)
an oscillation pattern of this same magnitude would be very significant. This could be achieved by running longer and/or building a larger detector
and/or using a more powerful reactor.

Another consideration is the energy resolution, or the minimum size of the bin in $E$. In order
to cleanly define the oscillation length in $L/E$, four data points per cycle are needed, or a resolution of 0.7 (0.33) m/MeV
for $\Delta m^2 = $ 0.9\, (1.9) eV$^2$. This corresponds to an energy resolution of 10\% (5\%). For $\Delta m^2 = 0.9$ eV$^2$, this 
resolution is less
stringent than that of the Bugey experiment by about 25\%.   For $\Delta m^2 =  1.9$ eV$^2$, the needed resolution is about double that in the Bugey 
experiment thus requiring a total 32 fold increase in counts relative to Bugey.

We estimated these numbers for 
an average energy neutrino. The result will approximately hold for smaller $L/E$ as can be seen in Fig.~\ref{fig2}. Focusing on an $L/E$ below 
5 m/MeV, one could combine 
two bins into one wider bin, reducing the error bars so that they become comparable to the other error bars, while retaining the requisite number of 
data points per oscillation. However, for the larger values of $L/E$ equal spacing in $E$ yields wider spacing in $L/E$, and the ideal spacing is not achieved.

There is an absolute maximum mass-squared difference that can be reached by this type of experiment. This is set by the physical size of the 
reactor. If one gets to a region where the length of a single oscillation is comparable to the dimensions of the reactor core, then neutrinos 
from the back of the reactor are incoherent with the neutrinos from the front of the reactor. Using a scale factor for the reactor 
core of 3 m and an average energy of 3.5 MeV, we find the 
maximum achievable mass-squared difference to be roughly 3 eV$^2$. This number depends on the shape of the reactor core, the location of 
the detector with respect to the orientation of the cylindrical core, and the power distribution within the core. A straightforward calculation 
for a given
experiment is necessary to get more than a crude estimate. Unfortunately, this number is smaller than the 10 eV$^2$ that is the lower end of a
presently allowed region found in Ref.~\cite{sore}. To reach this maximum sensitivity, an energy resolution of 3\% is required.

To determine the lower limit on the measurable mass-squared difference, we require that the oscillation phase reach at least $\phi= \pi/4$.  From this phase, one could determine the oscillation parameters without knowledge of the absolute flux.  
This gives a lower limit of 
$\Delta m^2 = 0.05$ eV$^2$ for $L=15$ m.

The measurement of $\Delta m^2$ is insensitive to the absolute normalization of the data; the oscillation length derived from the 
harmonic oscillations in the data determines $\Delta m^2$. The amplitude of these oscillations determines  directly $\sin^2 (2\,\theta_{14})$ which is also 
independent of the absolute flux if several oscillation cycles are measured. How this works out in the data analysis can be seen in Fig.~\ref{fig2}. 
For small values of $L/E$, the data are coherent and 
thus the peak is quite near ${\mathcal P}_{ee}=1$. The uncertainty in the norm of the data will necessarily be 
sufficient to allow the data to be uniformly adjusted such that the fit curve will have the peak for small $L/E$ also quite near ${\mathcal P}_{ee}=1$,
the required physical value. A great advantage of the proposed experiment is that the normalization of the data, usually the largest systematic error 
in a neutrino oscillation experiment, is nearly irrelevant.

The above utilized a baseline $L=15$ m. Doubling the flux by moving to $L=11$ m would be attractive. We repeat the analysis for 
this value of $L$. The maximum mass-squared difference which can be probed remains around 3 eV$^2$. The minimum and maximum value of $L/E$ shift 
to 2.2 and 10.3 m/MeV, respectively. The midpoint
for $L/E$ moves to 5.1 m/MeV, and the measurement would cover 4 (7) cycles for $\Delta m^2 = $ 0.9 (1.9) eV$^2$. To reach 3 eV$^2$, a resolution 
of 4\% in the energy would be required. By lowering the maximum value of $L/E$ the minimum sensitivity is raised, here to 0.06 eV$^2$. 
Going to a smaller baseline increases the flux by $L^2$, but the size of the reactor, hence its power, would necessarily decrease by $L^3$. This produces 
a smaller number of cycles such that the number of energy bins decreases by an additional factor of $L$. The overall result is that the experiment, 
to lowest approximation, is independent of the size of the reactor. This is true if you are searching for the 
existence of a sterile neutrino. Since the accuracy of the measured mass-squared difference would be increased by observing additional cycles, 
the last factor of $L$ would not apply if the goal were a fixed error on $\Delta m^2$. The larger width in the energy binning also reduces the 
resolution needed. For $L=11$ m, the required resolution
increases to 6 (13)\% 
for $\Delta m^2 = $ 0.9\, (1.9) eV$^2$. For a discovery experiment, 
a small research reactor with a small detector sitting very near the core is an interesting option to consider. The detector might even be 
wrapped around the core to increase the count rate.

The argument does require that the technology for neutrino detection scales nicely as the size of the detector. 
The technical issues for doing this experiment and those involved with using short baseline neutrino detectors for 
nonproliferation monitoring  \cite{sblnd} are related.

We have shown that a very short baseline reactor experiment can be used to measure the mass-squared difference associated with a sterile neutrino should it lie in 
the range of less than a tenth of an eV$^2$ up to several eV$^2$ by measuring the oscillations in the data over a number of oscillation 
lengths. A Fourier transform analysis of such data would be an efficient way of extracting the oscillation frequencies; however, such methods are 
not fundamentally different from fitting 
oscillation parameters to the data as a function of the mass-squared difference $\Delta m^2$. The experiment also measures
$\sin^2(2\,\theta_{14})$ through 
the amplitude of the oscillations. We suggest that a small reactor with a very short baseline 
be investigated. The experiment is not sensitive to the absolute normalization of the data and thus holds the possibility of being more accurate than 
alternatives.

The MiniBooNE experiment \cite{minib} will soon confirm or contradict the LSND experiment. The experiment proposed here could play a significant role 
independently of that outcome. Should MiniBooNE leave the situation ambiguous, the proposed experiment could provide a cost effective, accurate, 
and independent way to resolve the situation. Should MiniBooNE confirm the existence of a sterile neutrino(s), then this experiment might
provide a more accurate measurement of the mass-squared difference and of $\theta_{14}$. Should MiniBooNE not see evidence for a sterile 
neutrino, it would be setting upper limits on the mixing angles. This proposed experiment might then be a way of further looking for a sterile neutrino
in this range or, should a null result be found, further reducing the allowed value of $\theta_{14}$.

\acknowledgments {This work is supported in part by a grant from the US Department of Energy DE-FG02-96ER40975. }

%\bibliography{mass_b}

%\end{document}

\end{document}